\documentclass[a4paper,12pt,twoside]{article}
\usepackage[T1]{fontenc}
\usepackage[affil-it]{authblk}
\usepackage{amsmath,amssymb,amsthm,mathrsfs,mathtools,array,booktabs,ulem}
\usepackage{newtxtext,newtxmath,wrapfig,slashed,color,soul,enumitem}
\usepackage[margin=1in,marginparwidth=2cm]{geometry}
\usepackage{marginnote}
\usepackage[caption=false,singlelinecheck=false]{subfig}
\usepackage[scaled=0.92]{helvet}
\usepackage{setspace}
\usepackage{cite,placeins}
\usepackage[colorlinks, urlcolor=black, citecolor=blue, link
  color=black]{hyperref}

\usepackage{color}

\newcommand{\blue}[1]{\color{blue} #1 \color{black}}

\renewcommand{\d}{\ensuremath{\mathrm{d}}}
\newcommand{\modulus}[1]{\ensuremath{\big| #1 \big|}}

\allowdisplaybreaks

\begin{document}

\title{\bf Comments on ``Can quantum statistics help distinguish Dirac from
Majorana neutrinos?'' (arXiv 2402.05172 [hep-ph])}

\author[1]{C.~S.~Kim\thanks{cskim@yonsei.ac.kr}}
\author[2]{M.~V.~N.~Murthy\thanks{murthymundur@gmail.com}}
\author[3]{Dibyakrupa Sahoo\thanks{Dibyakrupa.Sahoo@fuw.edu.pl}}

\affil[1]{\normalsize Department of Physics and IPAP, Yonsei
  University, Seoul 03722, Korea}
\affil[2]{\normalsize The Institute of Mathematical Sciences (Retired), Taramani, Chennai 600113, India}
\affil[3]{\normalsize Institute of Theoretical Physics, Faculty of
  Physics, University of Warsaw,\newline ul.\ Pasteura 5, 02-093
  Warsaw, Poland}

\maketitle

\begin{abstract}
In a recent article~\cite{Akhmedov:2024}, the authors discuss the
question ``whether quantum statistics can help distinguish between
Dirac and Majorana neutrinos.'' The paper contains, among other
things, an unsubstantiated critique of the results derived in our
papers ~\cite{Kim:2021dyj} and \cite{Kim:2023iwz}. One of the
criticisms is related to results obtained using back-to-back
kinematics in the $B^0$ meson differential decay rate. We show that
the claim is wrong and point out how the correct result was obtained. 
The second criticism is related to the implementation of the
anti-symmetrization as dictated by quantum statistics for Majorana
neutrinos and antineutrinos \blue{(which are identical, by
definition). Any direct observation of the neutrinos, as done in
Ref.~\cite{Akhmedov:2024}, would project the neutrinos into
distinguishable helicity states, thus nullifying all observable
effects of quantum statistics.} They have missed the point that our
procedure holds when the neutrino-antineutrino remain undetected by
the detector. In the back-to-back kinematic configuration, one can
\textit{infer} the neutrino energies without directly detecting their
identities. This smartly ensures that the quantum statistical effects
are not erased. Their overriding assertion that our papers
\cite{Kim:2021dyj} and \cite{Kim:2023iwz} are incorrect fails to
recognize that in both Ref.~\cite{Kim:2021dyj} and \cite{Kim:2023iwz}
we also point out generic conditions under which the practical
Dirac-Majorana confusion theorem holds. ``Clearly there is no
confusion over confusion theorem.''
\end{abstract}

\section{Rebuttal of comments made in Sec.~3.1 of Ref.~\cite{Akhmedov:2024}}

Below we present our arguments highlighting our disagreement with the
incorrect statements in Ref.~\cite{Akhmedov:2024} in their Sec.3.1.
Their criticism is primarily centered around Eqs.~(48a) and (48b) of
our paper \cite{Kim:2021dyj} which give the differential decay rate
for the decay $B^0 \to \mu^- \, \mu^+ \, \nu_\mu \,
\overline{\nu}_\mu$ when the neutrino and antineutrino are
back-to-back in the rest frame of the parent meson $B^0$. Our work
basically aims at using the quantum statistics of identical particles
to distinguish between Dirac or Majorana nature of neutrinos in
general and using the above decay process in particular.

\subsection{The misunderstanding!}

The authors of Ref.~\cite{Akhmedov:2024} have clearly not followed our
step-by-step analysis of the back-to-back case which is correctly and
unambiguously described in Sec.IV.H of our paper~\cite{Kim:2021dyj}.
They simply state in Sec.~3.1 of their paper \cite{Akhmedov:2024} that
``the denominators of the left-hand sides of Eqs. (48a) and (48b)'' in
our paper \cite{Kim:2021dyj} ``erroneously contain $d \sin\theta$
instead of $d \cos\theta$'' without giving details but in a hand
waving manner. As explained in the next subsection, the phase space is
fixed by the final states of the process chosen. However, the
variables chosen to describe the differential decay rate are a matter
of choice once the kinematic frame is defined.\footnote{\blue{Changing
the variable of integration from $\sin\theta$ to $\cos\theta$ requires
change of the integrand as well so that the result after integration
remains the same. One can not simply change the variable of
integration while keeping the integrand unchanged.}}

The authors of Ref.~\cite{Akhmedov:2024} correctly state that the
back-to-back kinematics is ``fully characterized by just two physical
variables -- the muon energy $E_\mu$ and the angle $\theta$ between
the muon and neutrino directions.'' However, the article suggests that
 ``since the neutrinos are not observed, one must integrate over the
solid angle of the neutrino direction, which involves the integration
over $d \cos\theta$, not over $d \sin\theta$.'' If we start from the
general kinematics and correctly follow the change of variables
required to describe the back-to-back kinematics, we do find that one
gets $\d\sin\theta$ instead of $\d\cos\theta$. This important result
is explained in detail in our paper ~\cite{Kim:2021dyj}. We refer the
reader to see Sec.IV.F--H of our Ref.~\cite{Kim:2021dyj} in order to
get the correct understanding of the origin of $\d\sin\theta$ as
deduced from general kinematics to the back-to-back kinematics.

\subsection{Why a simple approach does not yield the correct decay rate}

The authors of\cite{Akhmedov:2024} suggest that ``One can consider the
problem in the back-to-back kinematics from the outset, without any
limiting procedures.'' However, such a simple  approach of imposing
the kinematics constraints artificially does not lead to the decay
rate but to some unphysical quantity. The correct approach is to start
with general formulation of the phase space and reduce it to the
desired kinematic situation as shown in in Sec.IV.F--H of our
paper~\cite{Kim:2021dyj}.

For the decay considered in Ref.~\cite{Kim:2021dyj} $B^0(p_B) \to
\mu^- (p_-) \, \mu^+ (p_+) \, \overline{\nu}_\mu (p_1) \,
\nu_\mu(p_2)$, the decay rate (considering all allowed kinematic
configurations) is given by,
\begin{equation}\label{eq:decay_rate}
\Gamma^{D/M} = \frac{1}{2\,m_B} \, \int \, \big| \mathscr{M}^{D/M}
(p_+, p_-, p_1, p_2) \big|^2 \, \d \textrm{LIPS} \; (2\pi)^4 \,
\delta^{(4)} \Big(p_B - p_+ - p_- - p_1 - p_2 \Big),
\end{equation}
where $\mathscr{M}^{D/M}(p_+, p_-, p_1, p_2)$ is the general amplitude
in Dirac ($D$) or Majorana ($M$) neutrino case, and the Lorentz
invariant phase space element is given by,
\begin{equation}
\d\textrm{LIPS} = \left(\frac{\d^3 \vec{p}_+}{(2\pi)^3 \, 2
E_+}\right) \, \left(\frac{\d^3 \vec{p}_-}{(2\pi)^3 \, 2 E_-}\right)
\, \left(\frac{\d^3 \vec{p}_1}{(2\pi)^3 \, 2 E_1}\right) \,
\left(\frac{\d^3 \vec{p}_2}{(2\pi)^3 \, 2 E_2}\right),
\end{equation}
where $p_i = (E_i,\vec{p}_i)$ with $i\in \{+,-,1,2\}$.
Since the mass dimension of the amplitude square in
our case is $-2$, Eq.~\eqref{eq:decay_rate} yields the correct mass dimension for the decay rate 
$\Gamma^{D/M}$,
\begin{equation}
    \textrm{mass dim} \big[ \Gamma^{D/M} \big] = 1.
\end{equation}

Following the suggestion of Ref.~\cite{Akhmedov:2024} to impose the
back-to-back kinematic condition (i.e.\ $\vec{p}_+ +\vec{p}_- =
\vec{0} = \vec{p}_1 + \vec{p}_2$) ``without any limiting procedures''
amounts to multiplying the right hand side of
Eq.~\eqref{eq:decay_rate} by the Dirac delta function $\delta^{(3)} \,
\left(\vec{p}_1 + \vec{p}_2\right)$. Thus, we get some quantity, say
$X^{D/M}$, given by

\begin{equation}\label{eq:XDM}
X^{D/M} = \frac{1}{2\,m_B} \, \int \, \big| \mathscr{M}^{D/M}(p_+,
p_-, p_1, p_2) \big|^2 \, \d \textrm{LIPS} \; (2\pi)^4 \, \delta^{(4)}
\Big(p_B - p_+ - p_- - p_1 - p_2 \Big) \, \delta^{(3)} \left(\vec{p}_1
+ \vec{p}_2\right).
\end{equation}
It is indeed true that the integration on right hand side is now much simpler and straightforward and it leads to
\begin{align}
X^{D/M} &=\frac{1}{32 \, (2\pi)^8 \, m_B} \int \big|
\mathscr{M}^{D/M}(\underbrace{p_+,p_-}_{\vec{p}_- =
-\vec{p}_+},\underbrace{p_1,p_2}_{\vec{p}_2 = - \vec{p}_1}) \big|^2
\left(\frac{\d^3 \vec{p}_+}{\modulus{\vec{p}_+}^2 + m_\mu^2}\right) \,
\left(\frac{\d^3 \vec{p}_1}{\modulus{\vec{p}_1}^2 + m_\nu^2}\right)
\nonumber\\*%
&\quad \times \delta \left(m_B - 2 \sqrt{\modulus{\vec{p}_+}^2 +
m_\mu^2} - 2 \sqrt{\modulus{\vec{p}_1}^2 + m_\nu^2} \right).
\label{eq:X-1}%
\end{align}
From usual analysis of solid angles we have, $\d^3 \vec{p}_+ =(2\pi)
\, E_\mu \, \sqrt{E_\mu^2 - m_\mu^2} \, \d E_\mu \, \d\cos\theta$, as
well as, $\d^3 \vec{p}_1 = (4\pi) \, E_\nu \, \sqrt{E_\nu^2 - m_\nu^2}
\, \d E_\nu$. Neglecting $m_\mu$ and $m_\nu$ dependence it is easy to
show that,
\begin{equation}\label{eq:X-2}%
X^{D/M} \approx \frac{1}{32 \, (2\pi)^6 \, m_B} \int \big|
\mathscr{M}^{D/M}(\underbrace{p_+,p_-}_{\vec{p}_- =
-\vec{p}_+},\underbrace{p_1,p_2}_{\vec{p}_2 = - \vec{p}_1}) \big|^2 \,
\d E_\mu \, \d\cos\theta,
\end{equation}
which indeed has $\d\cos\theta$ instead of $\d\sin\theta$ but the mass dimension of the quantity $X^{D/M}$ is 
\begin{equation}
\textrm{mass dim} \big[ X^{D/M} \big] = -2,
\end{equation}
due to the fact that $\delta^{(3)} \left(\vec{p}_1
+ \vec{p}_2\right)$ is a dimensionful quantity. 
Therefore $X^{D/M}$ is not the decay rate for the back-to-back kinematic configuration. On the other hand, our expressions Eq.~(48a) and (48b) in Ref.~\cite{Kim:2021dyj} have the  correct mass dimension $1$ for the corresponding decay rate in the back-to-back case. It 
should be obvious from the outset, since 
the Dirac delta-function is not a projection operator.

Therefore, any specific choice of kinematics has to follow from the general analysis by a suitable 
change of variables which appropriately defines the differential decay rate. 
We follow this step-by-step approach in \cite{Kim:2021dyj} to get the differential decay rate for back-to-back kinematic configuration. And this automatically leads us to $\d\sin\theta$ and not $\d\cos\theta$ in Eqs.~(48a) and (48b) in Ref.~\cite{Kim:2021dyj}. Further details on the analysis of back-to-back kinematics can also be found in Appendix A of Ref.~\cite{Kim:2023iwz}.

\subsection{Main conclusion of our papers \cite{Kim:2021dyj,Kim:2023iwz}}%

\blue{%
The authors of Ref.~\cite{Akhmedov:2024} add in their conclusion that
the main conclusion of our papers is dependent on the angular
distribution with respect to $\sin\theta$. We would like to emphasise
that the process $B^0 \to \mu^- \, \mu^+ \, \nu_\mu \,
\overline{\nu}_\mu$ and the choice of back-to-back kinematics
considered in \cite{Kim:2021dyj, Kim:2023iwz} are meant to serve as
illustrative examples of the many possibilities where the effect of
quantum statistics may be realised directly. Thus the main conclusion
of our papers is that quantum statistical effects can be used to pin
down the nature of the neutrino -- whether it is Majorana or Dirac
type.

In the context of processes allowed in the Standard Model, this
follows when
\begin{enumerate}
\item the direct and exchange terms are non-trivially
different\footnote{\blue{Due to this non-trivial difference between
direct and exchange terms in amplitude square, the phase space
integration for certain specific kinematic configuration, such as our
back-to-back case, need not vanish as well. Finding such kinematic
configuration and doing the relevant phase space integration depend on
the process under consideration. One can not make a process
independent observation at this level contrary to the goal of
Ref.~\cite{Akhmedov:2024}.}} (e.g.\ see Eqs.~(31) and (32) in
Ref.~\cite{Kim:2021dyj}), and%
\item the observable defined in terms of neutrino momenta is
accessible even when the neutrinos are not directly observed in the
final state.
\end{enumerate}
This is more clearly mentioned in Sec.2.2 and Fig.~1 of
\cite{Kim:2023iwz}, as well as in beginning of Sec.III of
\cite{Kim:2021dyj}. We stress that direct observation of the state of
neutrinos destroys the effect of quantum statistics since the
neutrinos in the final state are projected on to opposite helicity
states (and thus no longer remain indistinguishable) by observation.
Hence, any inference from qauntum statistical probe necessarily
requires no direct detection of the final neutrinos, as stated in
point 2 above. This is precisely achieved in the back-to-back
kinematic configuration where the energy-momentum information is
inferred purely by observing the muons leaving the antisymmetrization
effects of quantum statistics intact.

The differential decay rate in terms of $\sin\theta$ is the final
expression for the back-to-back configuration of the decay $B^0 \to
\mu^- \, \mu^+ \, \nu_\mu \, \overline{\nu}_\mu$. This just comes from
phase-space analysis alone and not from quantum statistical
considerations at all. We have clearly mentioned all the steps that
lead to the specific $\theta$ dependence of our result in
Ref.~\cite{Kim:2021dyj}. }

\section{Comments on Sec.~4 and ~5 of Ref.~\cite{Akhmedov:2024}}

In Sec.~4 of Ref.~\cite{Akhmedov:2024} the authors study the effect of
observing the active neutrinos via weak charged current (CC) and
neutral current (NC) processes in some neutrino detector (see their
Fig.~1b, 1c and Fig.~2). Their study suggests that when neutrinos get
detected, the difference between Dirac and Majorana neutrinos
vanishes. This is a well-known quantum measurement effect: if neutrino
and antineutrino can be distinguished by their detection, it destroys
the quantum indistinguishability of Majorana neutrino and
antineutrino. To study the effects of quantum statistics, one needs to
consider only Fig.~1a, without the quantum measurement effects that
come in Fig.~1b, 1c and 2. \blue{In Figs.~1b, 1c and 2 of
Ref.~\cite{Akhmedov:2024}, each of which is not a single Feynman
diagram but a combination of independent Feynman diagrams for
production and detection processes, the propagating neutrinos are not
virtual but real on-shell neutrino mass eigenstates which lead to
neutrino flavor oscillations. Such neutrinos even if they have
Majorana nature can not be considered as identical as one observes
their identity ($\nu$,  $\bar\nu$ or chirality) and position (left
detector and/or right detector which could be a few tens of meters
apart). It is very important that one takes into consideration whether
effects of measurement would adversely affect the quantum statistical
effects one is trying to probe.} As is shown in
Ref.~\cite{Kim:2023iwz} there is no difference between Dirac and
Majorana neutrino possibilities in $Z^{(*)} \to \nu \overline{\nu}$
decay mode, if one sums over the final spins of neutrinos. If one
wants to study spin dependent effects, one would need to detect the
neutrinos and that would have the measurement effect leading to no
difference between Dirac and Majorana possibilities, as the authors of
Ref.~\cite{Akhmedov:2024} have shown.\footnote{If the neutrino and
antineutrino interact with the detector by NC interaction, then their
spin states are still undetected and therefore the final spins must be
summed over. This, as we know, results in no difference between Dirac
and Majorana cases in $Z^{(*)} \to \nu \overline{\nu}$ decay in the
SM, see Ref.~\cite{Kim:2023iwz} and Sec.~4.2 of
Ref.~\cite{Akhmedov:2024}.}

This quantum measurement effect is analogous to the well-known fact
that the interference pattern in a double slit experiment is destroyed
if one could somehow know through which slit the photon has passed.
This is clearly highlighted in the last paragraph of Sec.~2.1 of
Ref.~\cite{Kim:2023iwz}, which says ``It should be noted that in this
work we discuss processes where the effect of measurements does not
destroy the identical nature of Majorana neutrino and antineutrino.
This is akin to putting the constraint that in a double-slit
experiment, meant to observe the interference of light, no measurement
should identify the slit through which the photon has passed." %
In summary, to correctly study signatures of quantum statistics it is
important that no measurements should affect the amplitude for the
process, or alter the predictions for the appropriate observable,
\textit{i.e.\ } quantum statistics requires absolutely identical
indistinguishable particles.

Therefore, the authors of Ref.~\cite{Akhmedov:2024} in their Sec.4 are
mistaken in wrongly suggesting that the quantum statistical
differences between Dirac and Majorana natures of active neutrinos
could be probed by detecting the active neutrinos and antineutrinos.
As explained above and in Ref.~\cite{Kim:2023iwz} such measurement
effects would alter the predictions for the very observable itself
which one is trying to study. Therefore, we disagree with the view of
the authors of Ref.~\cite{Akhmedov:2024} that quantum statistics is
not a useful tool to distinguish between Dirac and Majorana neutrinos.
In our works~\cite{Kim:2021dyj, Kim:2023iwz}, we smartly used the
back-to-back kinematics to deduce the energies of neutrinos, without
detecting their identities so as to avoid quantum measurement effect,
and correctly probe the quantum statistics.

\section{Comment on anti-symmerization requirement for Majoarana neutrino antineutrino pair and the massless limit}

When the neutrinos are not detected in the detector and their
polarisations are not directly deducible from the process itself, the
"chiral'' nature of neutrino or antineutrino is irrelevant. In any
case massive neutrinos are not chiral eigenstates. The full
calculation then involves sum over the spins of the unobserved
neutrinos and antineutrinos. This automatically take into account any
$m_\nu$ dependence. This is preciesly what is done in other processes
involving identical fermions, Moller scattering for example.

Besides, if one considers neutrino and antineutrino as chiral
fermions, then in the massless limit the Majorana condition is
mathematically impossible to satisfy (see Appendix B of
\cite{Kim:2023iwz}). Strictly massless neutrino and antineutrino, are
Weyl fermions and are distinct from one another for all purposes. Thus
a smooth massless limit for Majorana neutrino is a red herring.  The
effect of quantum statistics, an intrinsic property, can not depend on
a dimensional parameter! This is the reason why we consider only
non-zero mass for the neutrinos in Refs.~\cite{Kim:2021dyj} and
\cite{Kim:2023iwz}.

For more clarifications on the issues (i) one-to-one correspondence
between Dirac and Majorana neutrinos in the massless limit and (ii)
amplitude anti-symmetrization for the pair of Majorana neutrinos of
the same flavor,  we urge the reader to see Sec.~5 of
Ref.~\cite{Kim:2023iwz}.

\section{Summary} 

To summarise, we have pointed out how the step-by-step
approach~\cite{Kim:2021dyj} automatically leads us to $\d\sin\theta$
and not $\d\cos\theta$ for differential decay rate in Eqs.~(48a) and
(48b) in Ref.~\cite{Kim:2021dyj}.

Furthermore, we again reiterate that when the final state neutrino and
antineutrino are directly detected there will be no difference between
Dirac and Majorana scenarios due to the well known quantum measurement
effect.  In our case, we are considering a scenario where there is no
direct detection of the neutrino and antineutrino. The momenta are
gleaned indirectly by a clever use of the back-to-back geometry which
does not destroy the identical nature of Majorana neutrino and
antineutrino.  Hence the effect of quantum statistics remains intact
and this is the focus of our study. In the process we have clearly
shown the conditions under which the ``confusion theorem" may be
evaded and bring out the effect of quantum statistics. For a detailed
explanation Sec.~2.2 and Fig.~1 of \cite{Kim:2023iwz} especially for
(A) the requirement that there should be non-trivial difference
between the squares of direct and exchange terms, (B) the need to
infer the energies or 3-momenta of the neutrinos instead of direct
detection of the neutrinos, and (C) when one considers beyond SM
interactions.

\blue{We conclude by emphasizing that the detection and identification
of final neutrinos would erase all quantum statistical differences
between Dirac and Majorana neutrinos. However, this does not preclude
existence of quantum statistical differences when one is not directly
detecting and identifying the neutrinos and antineutrinos. It only
emphasizes that study of quantum statistical effects requires careful
consideration of only those signatures which can be measured or probed
without directly detecting the neutrinos. Therefore, we can not agree
with the conclusion of authors of Ref.~\cite{Akhmedov:2024} that
``quantum statistics does not lead to any exceptions to the Practical
Dirac-Majorana Confusion Theorem.'' Their ``general proof'' of the
practical Dirac Majorana confusion theorem fails to consider such
possibilities and takes into account only the measurement effects. We
would like to encourage that quantum statistical probes be further
explored carefully avoiding the nuisance of measurement effects.

\paragraph{Note added:} The authors of Ref.~\cite{Akhmedov:2024} also
cite Ref.~\cite{Marquez:2023rpc} to erroneously claim that integration
over the angle $\theta$ between the muon and neutrino directions in
our back-to-back kinematic configuration for $B^0 \to \mu^- \, \mu^+
\, \nu_\mu \, \overline{\nu}_\mu$ would give no difference between
Dirac and Majorana neutrinos, whereas in \cite{Marquez:2023rpc} they
integrate over the angle $\phi$ between the two decay planes (planes
made by the di-neutrino and the di-muon sub-systems) which leads to no
difference between Dirac and Majorana cases. In this context we would
like to point out our comments in \cite{Kim:2023ohr} where we
highlight how \cite{Marquez:2023rpc} is considering a different
process than ours and how their analysis of back-to-back kinematics is
erroneous as they integrate over the angle $\phi$ between the two
planes instead of fixing it to $\phi=0$.}

\section*{Acknowledgments}

We thank Prof. Evgeny Akhmedov for sharing their manuscript with us before submission and for discussions which brought more clarity to some issues. The work of CSK is supported by NRF of Korea (NRF-2022R1A5A1030700 and NRF-2022R1I1A1A01055643). The work of DS is supported by the Polish National Science Centre
under the grant number DEC-2019/35/B/ST2/02008.

\end{document}